\documentclass{openjournal}

\usepackage[utf8]{inputenc}
\usepackage[english]{babel}

\DeclareGraphicsExtensions{.pdf,.png,.jpg,.eps}

\usepackage[table]{xcolor}
\definecolor{linkcolor}{rgb}{0.0,0.3,0.5}

\usepackage{amsmath}
\usepackage{tensind}
\tensordelimiter{?}

\usepackage{textgreek}
\usepackage{verbatim}
\usepackage[normalem]{ulem}
\usepackage{soul}
\usepackage{orcidlink}

\usepackage{natbib}

\usepackage{hyperref}
\hypersetup{
    unicode, 
    colorlinks=true,
    linkcolor=linkcolor,
    citecolor=linkcolor,
    filecolor=linkcolor,
    urlcolor=linkcolor,
}
\graphicspath{{./figs/}}

\begin{document}
\title{Correlation of gamma-ray and optical variability of BL~Lacertae during the period of high blazar activity}

\author{M.S. Butuzova\orcidlink{0000-0001-7307-2193}}
\email{mbutuzova@craocrimea.ru}
\affiliation{Crimean Astrophysical Observatory of RAS, Nauchny, Crimea, Russia}

\author{M.A. Gorbachev\orcidlink{0000-0001-5731-9264}}
\email{mgorbachev17@gmail.com}
\affiliation{Crimean Astrophysical Observatory of RAS,  Nauchny, Crimea, Russia}


\begin{abstract}
Investigating the correlation of flux variability in blazars across multiple spectral bands provides a key tool for constraining emission mechanisms and jet physical parameters. We study the gamma-ray/optical variability correlation of BL Lacertae on timescales of $\sim$1 day during September 2022, when the source was within the TESS field of view, and over a longer period (April 2022–September 2023) encompassing its $\gamma$-ray flare, which finishes the series of flares during the high-activity phase. Our analysis confirms the Synchrotron Self-Compton mechanism as the origin of the $\gamma$-ray emission, characterized by a zero time lag. For the longer interval, the Discrete Correlation Function (DCF) reveals a second statistically significant peak at the time lag of about 12 days. 
We demonstrate that this peak arises due to the presence of an additional flare in one of the bands, which accompanies a $\gamma$-ray and optical flares correlated at zero time lag. We interpret the orphan flare origin in terms of the propagation of an injected blob of high-energy electrons along a jet with a decaying magnetic field, accounting for electron radiative cooling.
\end{abstract}


\maketitle

\section{Introduction}
\label{sec:intro}

Ultrarelativistic electrons in the jets of active galactic nuclei produce radiation at frequencies from radio to X-rays due to the synchrotron mechanism and at higher frequencies through inverse Compton scattering (ICS).
The seed photons for the latter process can have different origins: jet's synchrotron photons (synchrotron self-Compton, SSC), cosmic microwave background, radiation from stars of the parent galaxy, or broad line region \cite{Dermer93, Sikora94}.
Additionally, more complex scenarios may operate within jets, such as proton synchrotron radiation \cite{Aharon00} or proton-initiated cascades \cite{Mannheim93, Mannheim92}.
Therefore, studying the correlation of blazar emission variability across different spectral bands is an effective method for constraining emission mechanisms and physical parameters of relativistic jets. 
Although radio emission is subject to absorption within the jet medium \cite{Pushk12}, optical emission originates from regions located in the immediate vicinity of the true jet base. 
Thus, analyzing the relationship between $\gamma$-ray and optical variability will help clarify the localization of the $\gamma$-ray emitting region: either near the true jet base \cite{Dermer93, Tav10} or in a region of the jet located downstream of the 43~GHz VLBI core \cite{Marscher08}.

In the $\gamma$-ray range, light curves of bright sources have been continuous since the start of Fermi-LAT operations and are well sampled, whereas optical data are typically more sparse. 
To increase the sampling density of the time series for subsequent correlation analysis of variability on day-long timescales, international monitoring campaigns are carried out, such as those for S5~0716+714 \cite{Larionov13}. Despite certain technical details \cite{But25}, TESS provides the opportunity to obtain almost continuous light curves of individual sources with a duration ranging from 27 days to one year \cite{But25JHEA, Guseva25, Raiteri21-S50716, Raiteri21M-S40954, Otero-Santos24, Tripathi26, Weaver20}.

Searching for correlations on short timescales is particularly promising for bright, highly variable blazars. For example, from September 2020 to December 2023, BL~Lacertae (BL~Lac) exhibited strong variability at a historically high flux level \cite{Sahakyan22, Shah24, Xia26}. In September 2022, the source was within the field of view of TESS (Sector 56), enabling the acquisition of an optical light curve with high temporal resolution and almost continuous for 27 days. 
During the TESS observations, several prominent peaks are present in the well-sampled $\gamma$-ray and optical light curves of BL~Lac, providing an opportunity to investigate the correlation between $\gamma$-ray and optical variability.

The TESS observation interval covered the period immediately preceding a strong $\gamma$-ray flare,comparable in flux to the previous outbursts during the episode of high activity of BL~Lac, and ending it. 
Therefore, to study the correlation between optical and $\gamma$-ray variability, we also used data from April 2022 to September 2023, including the quiescent state before the flare, the complex-structured flare itself, and the subsequent low-amplitude flares.
The data used are described in Section~2. In Section~3, we identify the correlation between gamma-ray and optical variability.
Section~4 presents the investigation into the origins of the second significant peak in the discrete correlation function (DCF), the analysis of parameters affecting the characteristics of this second peak and the effect of the time-series sampling quality. 
Section~5 contains a discussion and conclusions.

\section{Gamma-ray and optical data}
\label{sec:obs_data}

\begin{figure}[!h] 
	\centering
	\includegraphics[width=.9\textwidth]{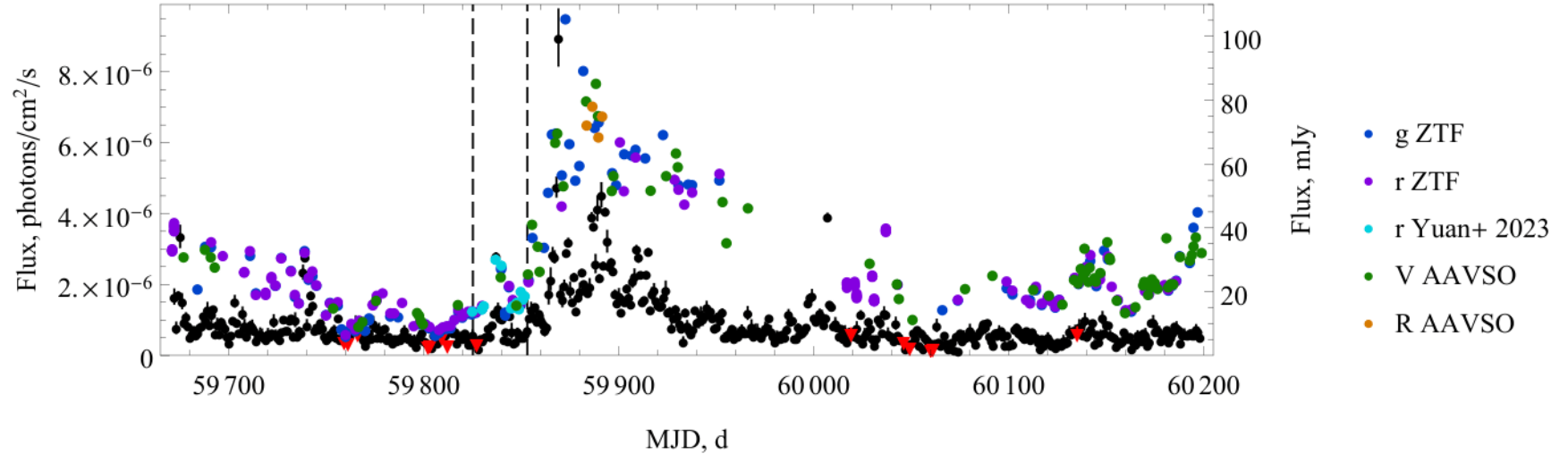}
	\caption{Long-term light curves of BL~Lac in the $\gamma$-ray and optical ranges, shown in black and colors, respectively. Red triangles mark $\gamma$-ray data with TS<9. The energy of the high-energy photons lies in the range 0.14$-$300 GeV. In the optical band, the light curve is constructed from ZTF, Yuan et al. \cite{Yuan}, and AAVSO data, all converted to the brightness in the ZTF g-band filter. Dashed lines denote the TESS observation period.}
	\label{FIG:1}
\end{figure}

In the $\gamma$-ray range, we used observational data from the Fermi-LAT space telescope, provided as part of the monitoring program\footnote{https://fermi.gsfc.nasa.gov/ssc/data/access/lat/msl\_lc/} for bright and transient sources whose flux has ever exceeded the established threshold of $2\cdot 10^{-6}$~photons/cm$^2$/s. 
For the monitored objects, light curves are available in the energy bands 0.3$-$1, 14$-$300, and 0.1$-$300 GeV, with weekly and daily time binning. In our analysis, we used the 0.1$-$300~GeV energy range with 1-day binning and Test Statistic TS>9 (Fig.~\ref{FIG:1}). The high brightness of the source throughout the entire considered period causes the almost absence of flux upper limits in the daily averaging $\gamma$-ray data.

\begin{figure}[h]
	\centering
	\includegraphics[width=.5\columnwidth]{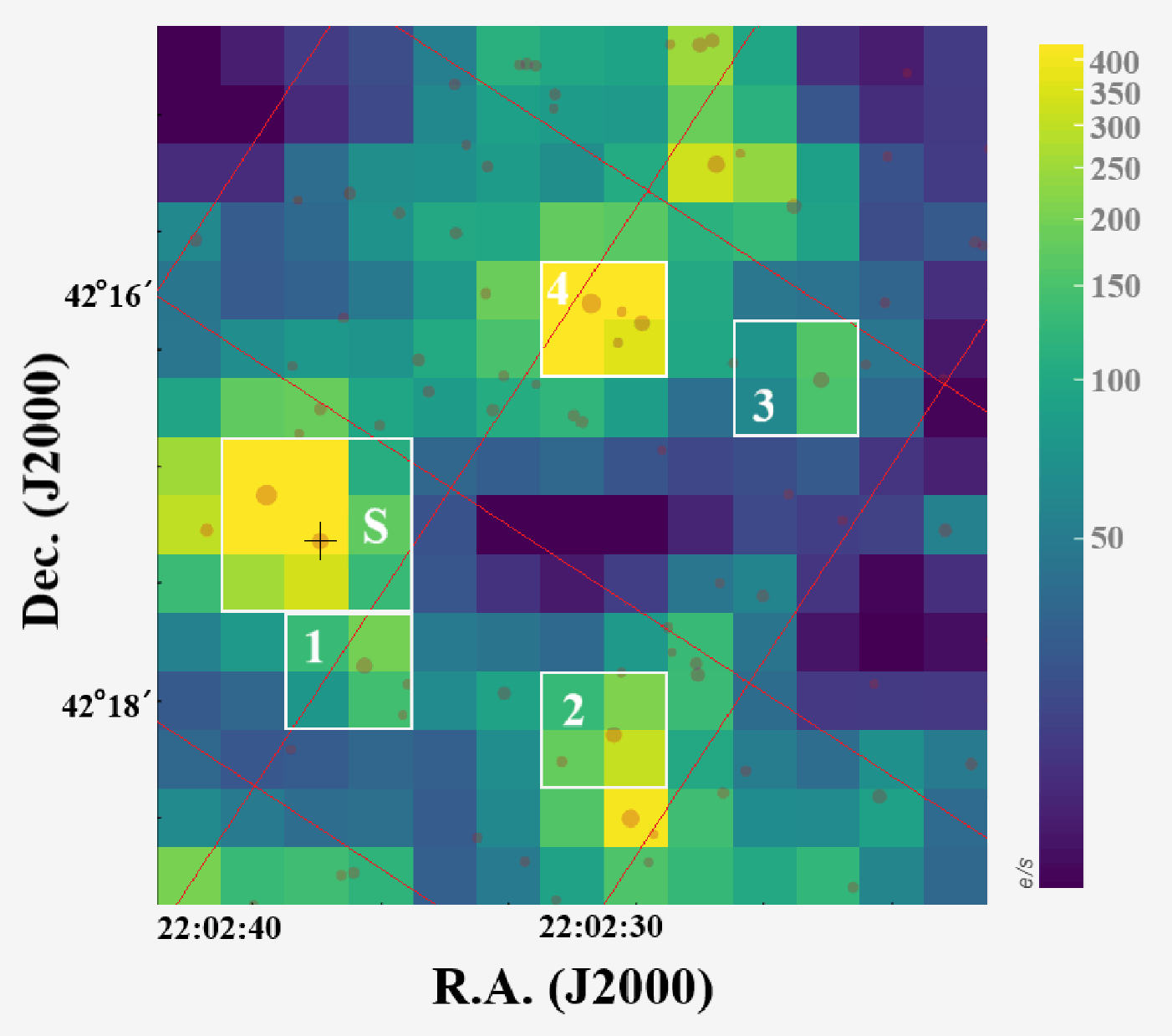}
	\caption{TESS full-frame image cut. BL~Lac is marked by a cross. The object and comparison star (1, 2, 3) apertures are displayed by white frames. Instrumental fluxes are color-coded.}
	\label{FIG:2}
\end{figure}

\begin{figure}[h]
	\centering
	\includegraphics[width=.9\textwidth]{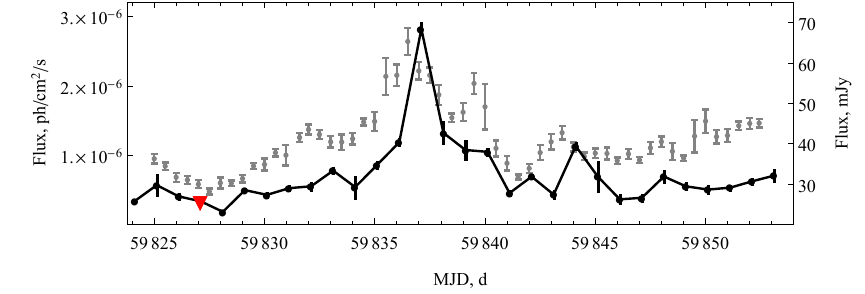}
	\caption{The short-term light curve of BL~Lac according to TESS observations and Fermi-LAT data in the range 0.1$-$300~GeV (gray and black, respectively). Red triangle marks $\gamma$-ray data with TS<9}.
	\label{FIG:3}
\end{figure}

We used two types of optical data. 
The first consists of TESS observations (MJD 59824$-$59852) obtained via aperture photometry of summed 10 consecutive cuts of full frame image following the algorithm described in \cite{But25JHEA} (Fig.~\ref{FIG:2}).
We adopted the aperture shown in Fig.~\ref{FIG:2} and subtracted the contribution of constant stars. 
Conversion of magnitudes to flux was performed using the TESS filter zeropoint ($\text{ZP}_\lambda=1.33 \cdot 10^{-9}$ erg/cm$^2$/s/\AA ~\cite{Rodrigo24, Rodrigo12}). The resulting TESS light curve, averaged over 0.5 days, is presented together with the $\gamma$-ray band light curve in Fig.~\ref{FIG:3}.

The second type comprises combined heterogeneous observations of BL~Lac over the interval MJD~59671$-$60200. For this purpose, we used ZTF monitoring data \cite{Bellm2019}
(in the g and r bands) along with the daily-averaged measurements from \cite{Yuan} and the AAVSO database. 
All data have been corrected for Galactic extinction with the coefficients from \cite{Schlafly2011}.
To construct densely sampled time series under conditions of insufficient data from simultaneous observations in two or more photometric bands, we combined the datasets as follows.

The ZTF g-band photometry was adopted as the reference light curve, being the largest homogeneous subset. The r-band data, after removing epochs coincident with g-band observations, were aligned to the reference light curve by applying a constant magnitude offset. This offset was determined by minimizing the sum of squared differences between the r-band measurements and the linear trend fitted to the reference light curve at the r-band observation epochs.
In the next step, the resulting combined dataset served as the new reference, to which the Yuan et al. \cite{Yuan} and AAVSO data were aligned using the same procedure. For the AAVSO data, measurements in different optical bands were first aligned to the reference light curve separately, and then appended to the combined dataset. The final light curve, converted from g-band magnitudes to flux, is presented in Fig.~\ref{FIG:1}.
We justify the validity of a linear shift of light curves for magnitudes in the Appendix~\ref{ap:app1}. 

The light curves (Figs.~\ref{FIG:1}, \ref{FIG:3}) reveal that the object exhibited variability during the TESS observations, with several flares of different amplitudes detected in both wavelength ranges. Around MJD~59870, a prominent gamma-ray flare is observed (Fig.~\ref{FIG:1}), comparable in amplitude to the preceding events, finished of the BL Lac's high-activity state.

\section{Correlation of gamma-ray and optical variability}

\begin{figure}[!h]
	\centering
	\includegraphics[width=.6\columnwidth]{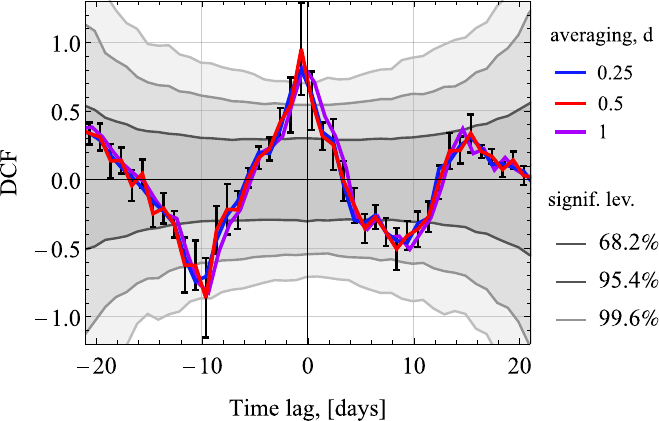}
	\caption{Discrete correlation function for the $\gamma$-ray and optical light curves over the period MJD~59824$-$59852 and its significance levels.}
	\label{FIG:4}
\end{figure}

To search for correlations that account for the time delay $\tau$ between variability in different spectral bands, we applied the Discrete Correlation Function (DCF) \cite{EdKrol}. To determine the lag on a one day time scale, we used TESS and Fermi-LAT data from the interval MJD~59824$-$59852. 
Note that the original TESS light curve has a time resolution of approximately 40 minutes and is oversampled for correlation analysis on daily timescales. 
Therefore, we use the averaged light curve. 
This filters out high-frequency noise, retaining only the signal associated with processes on the timescale under investigation. 
An averaging interval of 0.5 days allows, first, the detection of sub-daily delays to be preserved. 
Second, the shape of flares with durations $<1$ day is better preserved than with a larger averaging interval. 
The calculated DCFs for different intervals of averaging show peak near zero time lag (Fig.~\ref{FIG:4}).

To estimate significance levels, for each light curve the corresponding parametric time-series model was derived using Wolfram Research Mathematica and 10,000 simulated data series were generated based on this model. 
The simulations corresponding to the two different observed time series were paired, and the DCF was computed for each pair. 
Using the resulting 10,000 DCF realizations, for each value of the time delay $\tau$ we determined the intervals containing 68.2\%, 95.4\%, and 99.6\% of the correlation coefficient values; connecting these intervals yielded the significance levels corresponding to 1, 2, and 3$\sigma$, respectively (Fig.~\ref{FIG:4}).

The DCF shown in Fig.~\ref{FIG:4} exhibits a statistically significant maximum at a time delay of the optical variability relative to the $\gamma$-ray one of $-0.1\pm0.5$~days (for averaging over 0.5~d). 
The uncertainty in the lag determination was estimated by calculating the absolute deviation from the zero-lag axis of the peak position in the discrete autocorrelation function for each of the two time series.  
The larger value was adopted as the final uncertainty. In this case, the uncertainty was derived from the $\gamma$-ray data.

In the $\gamma$-ray band, the emission is produced via inverse Compton scattering, whereas in the optical band it arises from the synchrotron mechanism. 
The strong correlation between variability in these two bands suggests the operation of synchrotron self-Compton scattering, when relativistic electrons in the jet scatter their own synchrotron photons into the $\gamma$-ray range.

\begin{figure}[!h]
	\centering
	\includegraphics[width=.6\columnwidth]{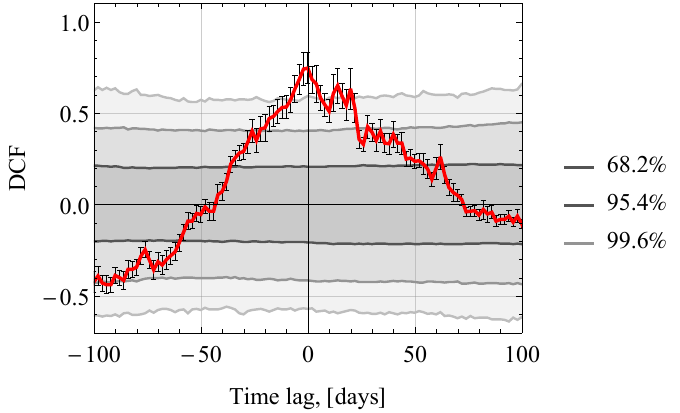}
	\caption{Discrete correlation function for the $\gamma$-ray and optical light curves over the period MJD~59672$-$60199 and its significance levels.}
	\label{FIG:5}
\end{figure}

The DCF calculated from long-term data (MJD~59671 $-$ 60200), with significance levels constructed using the algorithm described above, exhibits two statistically significant peaks (Fig.~\ref{FIG:5}). 
The first peak, occurring at a time lag close to zero, is less sharp than the peak observed in the DCF for the short-term data. 
Therefore, to find the exact peak position, the data points within the lag range $\tau = - 6$ to 6~days (inclusive) were fitted by a Gaussian function, yielding a maximum at $- 0.83\pm 1.06$~days.
The uncertainty in $\tau$ was estimated using the same procedure as in the previous case. The largest value was derived from the optical-band data. 
The second peak in the DCF occurs about 12~days and  has a long statistically significant tail of up to $\tau \approx 20$~days.

\section{Identifying the causes of the two-peak DCF profile}
\subsection{Light curve simulations\label{lcm}}

To investigate the origin of the secondary peak in the DCF, we examined the long-term light curves. Figure~\ref{FIG:1} shows that frequent flares of varying amplitudes occurred repeatedly in both bands. 
Consequently, the strong correlation may be spurious, arising primarily from the blazar’s heightened and coincidental flaring activity in both bands rather than indicating a direct physical connection.
In contrast to the Gaussian process model used to determine significance levels, we generated 10,000 pairs of light curves of the same duration, assuming that flares in each pair occur independently at random times and are superimposed on a steady signal.
The flare profile was set as
\begin{equation}
    F(t)=
    \begin{cases}
  a \cdot \exp[{b_r}(t-t_F)],   \text{~~~for ~} t<t_F \\
  a \cdot \exp[-{b_d}(t-t_F)],  \text{~for ~} t>t_F,
\end{cases}
\label{eq1}
\end{equation}
where $a$ is the amplitude, $b_r$ and $b_d$ are the parameters that govern the rise and decline rate, respectively, $t_F$ is the position of the flare peak.

The simulations were performed under the constraint that no flares occur during the first third of the time interval.
For each pair of light curves, the number of flares was identical in both series and was treated as a random variable uniformly distributed between 7 and 15 (to mimic high flaring activity). 
The same sequence numbers of flares in two series have equal $a$ and $b_r=b_d$.
These parameters are also set as the realization of a uniformly distributed random variable in a certain range of values.

\begin{figure}[!h]
	\centering
	\includegraphics[width=.8\textwidth]{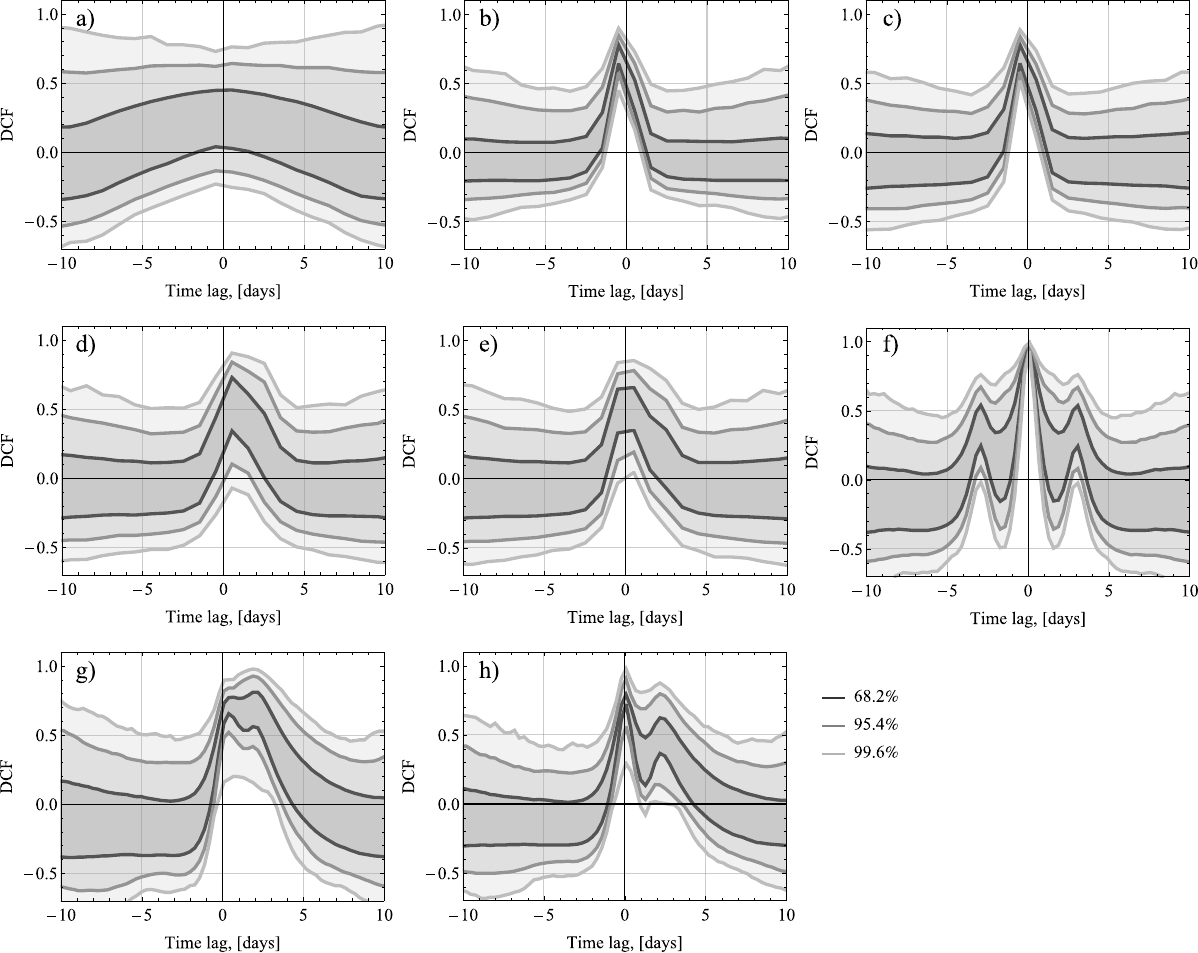}
	\caption{Intervals containing 68.2\%, 95.4\% and 99.6\% of DCF points calculated under different assumptions:
    a) independently occurring flares during the last two thirds of the considered time interval;
    b) correlated with zero time lag flares, having in
optical series more slow decay. The number of flares is a random variable in the range from 3 to 6;
    c) Same as in panel b), but with the number of flares ranging from 7 to 15.
    d) correlated flares with identical parameters, whose time lags follow a mixture of two normal distributions with a common variance of $\sigma^2=0.5$, component medians at 0 and 2, and mixing weights of 60\% and 40\%, respectively;
    e) correlated flares with identical parameters, whose time lags were randomly assigned to 0 in two-thirds of flare number and to 2 in one-third;
    f) simultaneous flares with a two-component peak structure, with sub-peaks separated by 2 t.u. The amplitude of the first sub-flare is twice that of the second;
    g) pairs of two-component $\gamma$-ray flare corresponding to a single optical flare (Case A); 
    h) pairs of two-component optical flare corresponding to a single $\gamma$-ray flare (Case B).
    }
	\label{FIG:6}
\end{figure}

The synthesized light curves spanned a duration of 25 time units (t.u.). The sampling intervals for the $\gamma$-ray and optical series were 0.1 and 0.05 t.u., respectively. 
The $\gamma$-ray and optical series were offset in time, starting at $t = 0$ and $t = 0.03$, respectively.
For each pair of simulated light curves, the DCF was calculated and confidence intervals containing 68.2\%, 95.4\% and 99.6\% of the values were determined at each time lag. As shown in Fig.~\ref{FIG:6}a, the 99.6\% interval exhibits predominantly weak correlations, with the maximum occurring at zero delay and reaching nearly 0.5.
The upper and lower boundaries of the 99.6\% interval gradually and symmetrically decrease  in two directions from $\tau-0$. This 99.6\% interval does not match the observed DCF, which compels us to consider alternative scenarios in our simulations.

Figure~\ref{FIG:1} shows a prominent flare around MJD~59860 $-$ 59880, detected in both bands. Notably, the decline for the optical flare was more prolonged than that in the high-energy range. 
Therefore, we modeled a scenario with zero-lag correlated flares in both bands, but with a smoother decline assigned to the optical-band light curve, implemented as $b_d=d_r/k$.
We obtained $k$ and $b_r$ as realizations of a random variable uniformly distributed in the range $[0.05,0.7]$ and $[1.5,3]$, respectively.
The duration and sampling of the light curves were kept constant in all simulated scenarios. In each simulated pair, both series contain the same number of simultaneous flares. Flares having the same sequence number in the two series were assigned identical amplitudes and $b_r$ values.
The confidence intervals containing 68.2\%, 95.4\% and 99.6\% of the DCF values, computed from 10,000 pairs of simulated light curves with 3--6 and 7--15 flares are shown in Fig.~\ref{FIG:6}b and \ref{FIG:6}c, respectively.
Intervals with different significance levels exhibit a peak with a small negative delay, the height of which can reach large values of the correlation coefficient. 
On both sides of the peak, the DCF declines rapidly and symmetrically; notably, the number of flares in the simulated time interval has no significant effect on the interval shape. The 99.6\% confidence intervals (Fig.~\ref{FIG:6}b, c) do not reproduce the shape of the observed DCF.

To reproduce a more gradual decline toward positive lags in the 99.6\% confidence interval -- and, ideally, a secondary peak -- we modeled pairs of light curves containing the same number of flares (randomly chosen from 5 to 15 for each pair). 
Flares with the same sequence numbers have the equal amplitudes, rise and decay rates, and the peak of the flare in the simulated optical light curve occurs with some delay relative to the $\gamma$-ray one. 
The peak of each flare in the simulated optical light curve was delayed relative to its $\gamma$-ray counterpart. 
This delay was set as a random variable, which is distributed over the mixture of two normal distributions with the same variance of 0.5 and component medians at 0 and 2, and mixing weights of 60\% and 40\%, respectively (Fig.~\ref{FIG:6}d), or taken values are randomly assigned to 0 in two thirds of the flare number and 2 in the remaining cases (Fig.~\ref{FIG:6}e). 
Both scenarios yield similar 99.6\% confidence intervals for the DCF, exhibiting a more gradual decline of the correlation coefficient toward positive lags. However, even at the peak ($\tau=0$), the DCF peak values are lower than in the previous cases.

Then, we modeled several (from 3 to 7 for an individual pair of light curves) simultaneous flares with two sub-peaks separated by 2 u.t. in both series. The amplitude of the first sub-peak was set to twice that of the second. This resulted in the shape of the 99.6\% confidence interval in the DCF with a pronounced central peak at $\tau=0$ and lower symmetric peaks on both sides (Fig.~\ref{FIG:6}f). This case also does not correspond to the DCF constructed from observational data.

Our next modeling assumption was that the number of flares in the two data series differed. First, we considered Case A, where a single optical flare corresponded to a double flare in the $\gamma$-ray band. The sub-peaks of the $\gamma$-ray flare were separated by 2~t.u. The rise and decay rates were set equal ($b_r=b_d$), with values ranging randomly from 1.5 to 3. The amplitude of the second sub-peak was set to twice that of the first, and its peak position coincided with the peak of the optical flare. The optical flare was characterized by $b_r$ equal to that of its $\gamma$-ray counterpart, and $b_d=b_r/5$. Based on 10,000 simulated pairs of light curves, a confidence interval containing 99.6\% of the DCF values was determined at each lag $\tau$. This interval exhibits a two-peak structure (Fig.~\ref{FIG:6}g), with one peak at $\tau=0$ and the second at positive lags. It is consistent with the observational data.

In the two-peak 99.6\% interval form, it is difficult to fit a single-peak separate DCF, unlike in the previous case. Therefore, one $\gamma$-ray flare corresponding to the two-peaked optical flare are most likely reproduce the two-peak DCF shape constructed from long-term $\gamma$-ray and optical observations.

Second, we considered Case~B, describing a double-peaked flare in the optical band. The first sub-peak had a simultaneous counterpart in the $\gamma$-ray series, with both peaks characterized by $b_r=b_d$. The second sub-peak occurred after a delay of $\Delta t=2$~t.u. and was assigned half the amplitude of the first, along with a more prolonged decline phase ($b_d=b_r/5$). By setting the number of $\gamma$-ray flares per light curve to 2 -- 4, we generated 10,000 simulated pairs; the resulting 99.6\% confidence interval for DCFs is shown in Fig.~\ref{FIG:6}h. Unlike in the previous case, it is difficult to match a single-peaked DCF to the two-peaked 99.6\% confidence interval. Therefore, a scenario in which a single $\gamma$-ray flare corresponds to a double-peaked optical flare is most likely to reproduce the two-peak DCF shape constructed from long-term $\gamma$-ray and optical observations.

\subsection{The Influence of synthetic flare parameters on the DCF shape}

We examine the DCF morphology for individual flare events corresponding to Cases A and B (Fig.~\ref{FIG:6}g and \ref{FIG:6}h, respectively).

For Case~A two-peaked flare was introduced in the model light curve associated with the gamma-ray band: the first sub-peak occurred at time $t_1$, and the second with a delay of $\Delta t$. 
The second sub-peak of $\gamma$-ray flare was aligned with the single peak in the model optical light curve.
The temporal sampling of the $\gamma$-ray and optical light curves was set to 0.1 and 0.05 t.u., respectively. The first points of the $\gamma$-ray and optical series were assigned to $t=0$ and $t=0.03$, respectively.

\begin{figure}[!h]
	\centering
	\includegraphics[width=.85\textwidth]{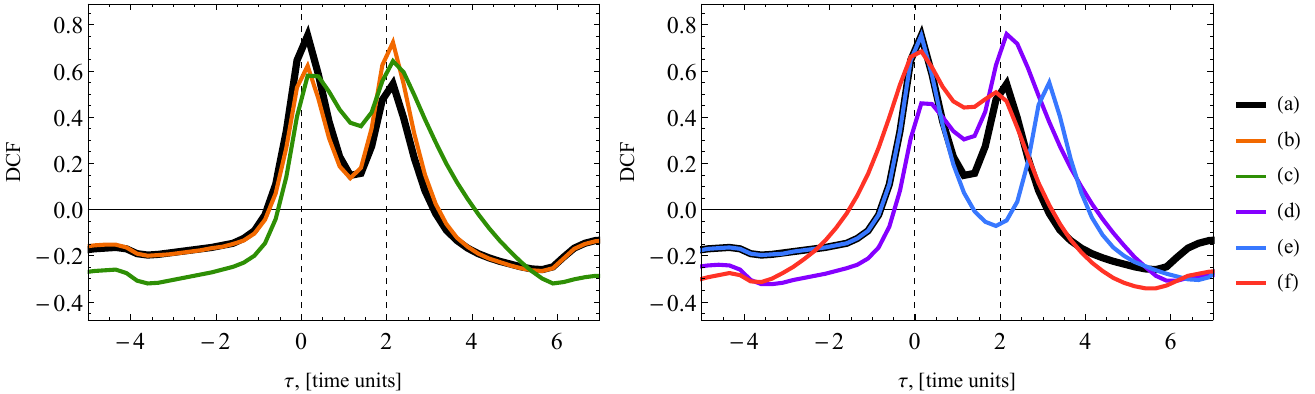}
	\caption{DCFs for model flares of Case~A with various parameters. In both panels, the DCF calculated using the parameters listed in column (a) is shown in black. The legend corresponds to the parameters provided in the respective columns of Table~\ref{tab1}.}
	\label{FIG:7}
\end{figure}

Following Equation (\ref{eq1}), we specified sub-peaks of varying amplitudes in the $\gamma$-ray modeled series and a more gradual flare decline in the optical ligth curve (see Table~\ref{tab1}). 
By varying the parameters $\Delta t$, the $\gamma$-ray sub-peak amplitude ratio $a_1/a_2$ and the optical decline rate, we obtained the corresponding DCFs that also exhibit the two-peaked structure, although with certain modifications depending on the parameter values (Fig.~\ref{FIG:7}).

Thus, increasing the amplitude of the first $\gamma$-ray sub-peak and prolonging the decline phase of the optical flare (i.e., decreasing the parameter $b_d$) leads to a reduction of the peak value at $\tau=0$ and an enhancement of the secondary peak of DCF. 
Decreasing the parameters $b_r$ and $b_d$ results in more gradual peaks in the DCF. 
The separation between the two DCF peaks equals the time interval between the $\gamma$-ray sub-peaks.

\begingroup 
    \setlength{\tabcolsep}{10pt} 
    \renewcommand{\arraystretch}{1.5} 
    \setlength\extrarowheight{2pt}
    \begin{table*}
        \centering
        \caption{Parameters of two-peaked $\gamma$-ray flare and correlated with it optical flare used in our simulations of case~A. For ease of comparison, parameter values in each column that differ from those in column (a) are highlighted in bold.}
        \label{tab1}
        \begin{tabular}{|c|cccccc|}  
            \hline \hline     
            \textbf{Parameters} \textbackslash \textbf{Cases} & (a) & (b) & (c) & (d) & (e) & (f) \\ 
            \hline
            $\gamma$-flares & ~ & ~ & ~ & ~ & ~ & ~ \\ 
            $t_1$ & 4 & 4 & 4 & 4 & 4 & 4 \\
            $\Delta t$ & 2 & 2 & 2 & 2 & \textbf{3} & 2 \\
            $a_1$ & $a_2/1.5$ & $\boldsymbol{a_2}$ & $a_2/1.5$ & $\boldsymbol{a_2}$ & $a_2/1.5$ & $a_2/1.5$ \\
            $a_2$ & 3 & 3 & 3 & 3 & 3 & 3 \\
            $b$ & 4 & 4 & 4 & 4 & 4 & 4 \\
            \hline
            optical-flare & ~ & ~ & ~ & ~ & ~ & ~ \\ 
            $a$ & 3.7 & 3.7 & 3.7 & 3.7 & 3.7 & 3.7 \\
            $b_r$ & $b$ & $b$ & $b$ & $b$ & $b$ & $\boldsymbol{b/4}$ \\
            $b_d$ & $b_r/2$ & $b_r/2$ & $\boldsymbol{b_r/6}$ & $\boldsymbol{b_r/6}$ & $b_r/2$ & $\boldsymbol{b/4}$ \\
            \hline \hline  
        \end{tabular}  
    \end{table*}
\endgroup

\begin{figure}[!h]
	\centering
	\includegraphics[width=.85\textwidth]{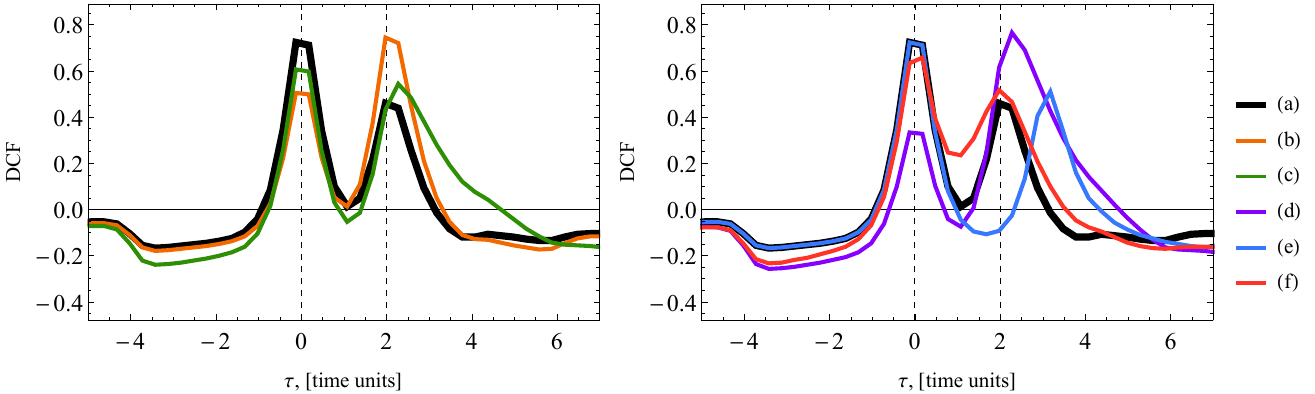}
	\caption{DCFs for model flares of Case~B with various parameters. In both panels, the DCF calculated using the parameters listed in column (a) is shown in black. The legend corresponds to the parameters provided in the respective columns of Table~\ref{tab2}.}
	\label{FIG:8}
\end{figure}

For case B, while maintaining the temporal characteristics of the model time series described above, we simulated a two-component optical flare. 
The first sub-flare was simultaneously with the model $\gamma$-flare and occurred at time $t_1$, while the second sub-flare followed after a delay $\Delta t$. 
The second sub-flare amplitude is 2 times less than for the first one. 
The DCF computed from these data is shown as a black line in the plots of Fig.~\ref{FIG:8}. 
By varying the flare parameters (see Table~\ref{tab2}), we concluded that the ratio of sub-flare amplitudes is the dominant factor determining the relative height of the second peak in the DCF: the smaller the ratio $a_1/a_2$, the more prominent the peak. 
Furthermore, increasing the duration of the decay (or rise) phase of the second optical sub-flare results in a more gradual decline (or growth) of the second DCF peak.

\begingroup 
    \setlength{\tabcolsep}{10pt} 
    \renewcommand{\arraystretch}{1.5} 
    \setlength\extrarowheight{2pt}
    \begin{table*}
        \centering
        \caption{Parameters of $\gamma$-ray flare and correlated with it two-peaked optical flare used in our simulations of case~B. For ease of comparison, parameter values in each column that differ from those in column (a) are highlighted in bold.}
        \label{tab2}
        \begin{tabular}{|c|cccccc|}  
            \hline \hline     
            \textbf{Parameters} \textbackslash \textbf{Cases} & (a) & (b) & (c) & (d) & (e) & (f) \\ 
            \hline
            $\gamma$-flare & ~ & ~ & ~ & ~ & ~ & ~ \\ 
            $a$ & 3.7 & 3.7 & 3.7 & 3.7 & 3.7 & 3.7 \\
            $b$ & 4 & 4 & 4 & 4 & 4 & 4 \\
            \hline
            optical-flare & ~ & ~ & ~ & ~ & ~ & ~ \\ 
            $t_1$ & 4 & 4 & 4 & 4 & 4 & 4  \\
            $\Delta t$ & 2 & 2 & 2 & 2 & \textbf{3} & 2 \\
            $a_1$ & 3 & 3 & 3 & 3 & 3 & 3 \\
            $a_2$ & $a_1/2$ & $\mathbf{a_1}$ & $a_1/2$ & $\mathbf{a_1}$ & $a_1/2$ & $a_1/2$ \\
            $b_{1,\,r}$ & $b$ & $b$ & $b$ & $b$ & $b$ & $b$ \\
            $b_{1,\,d}$ & $b$ & $b$ & $b$ & $b$ & $b$ & $b$ \\
            $b_{2,\,r}$ & $b$ & $b$ & $b$ & $b$ & $b$ & $\mathbf{b/4}$ \\
            $b_{2,\,d}$ & $b/2$ & $b/2$ & $\mathbf{b/6}$ & $\mathbf{b/6}$ & $b/2$ & $\mathbf{b/4}$ \\
            \hline \hline  
        \end{tabular}  
    \end{table*}
\endgroup

\subsection{The Influence of the irregularity of the data series}

\begin{figure}[!h]
	\centering
	\includegraphics[width=.85\textwidth]{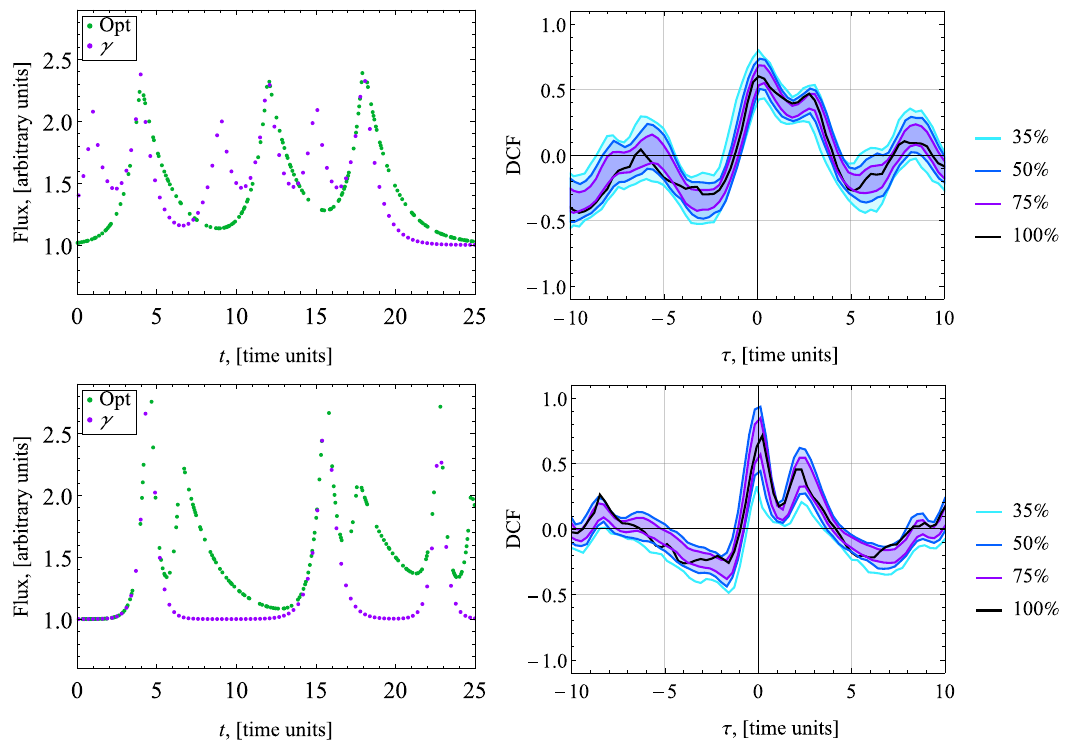}
	\caption{Model light curves (left panels) and the corresponding 99.6\% DCF confidence intervals (right panels) for cases A and B (upper and lower panels, respectively). The DCFs computed from the uniform original model light curves are shown by black lines.}
	\label{FIG:9}
\end{figure}

The simulation was performed for two uniform and densely sampled time series. This corresponds to uniform sampling of observational data in the $\gamma$-range, whereas in the optical range the time interval between data points ranges from 1 minute to 50.7 days. In 90\% of cases, this interval falls within 3.49 days and in 98\% of cases, within 7.26 days.
To investigate the effect of irregular sampling in the time series, we synthesized $\gamma$-ray and optical light curves under the assumption of three correlated flare events. We then generated 1000 light curves, each comprising 75\% of the data points from the initial optical light curve, randomly selected in each iteration. Using these series in combination with the model $\gamma$-ray light curve, we computed 1000 DCFs, from which we derived a 99.6\% confidence interval.
For further calculations, 50 and 30\% of the points of the initial optical light curve were selected in the same way. In the latter case, we additionally modified the $\gamma$-ray light curve by taking every third point from the initial one for the DCF computations. The results for cases A and B are presented in Fig.~\ref{FIG:9}.
It can be seen that even under significant deterioration in the discretization of both time series, the two-peaked shape of the DCF is preserved. This confirms the statistical reliability of our interpretation of the DCF derived from the observational data.

\section{Discussion and conclusions}
\label{sec:concl}

During the first 18 months of Fermi-LAT observations of BL Lacertae, the object was faint in the $\gamma$-ray band and exhibited flux variability uncorrelated with its optical variability \cite{Abdo11BLLac}. The spectral energy distribution (SED) can be fitted with several models, but an SSC model augmented by an external Compton component best accounts for the observed variability and satisfies the equipartition condition. During the high-activity period of BL Lacertae in 2011--2012, a $\gamma$-ray/optical flux correlation was observed, with a DCF peak of 0.6 and zero time lag, supporting a simple leptonic SSC model \cite{Rait13}. Raiteri et al. \cite{Rait13} note that the amplitude of the gamma-ray flux variability is approximately equal to the square of the optical variability amplitude, as predicted by the SSC theory (e.g., \cite{Konigl81}). The relatively low correlation coefficient might be attributed to both data gaps and a more complex structure of the optical flares, possibly arising from substructures within the optical emission region \cite{Narayan12}, some of which do not produce gamma-ray emission \cite{Rait13}. For example, in analysis of the multiwavelength SED during the submillimeter flare of October--November 2023, the Mondal et al. \cite{Mondal25} argue for the presence of two synchrotron components: one producing emission up to submillimeter wavelengths, and the other in the optical band. These components correspond to two distinct electron distributions. Additional evidence for the SSC scenario is provided by the analysis of polarization in the optical and X-ray ranges  \cite{Agudo25}.

The zero-lag correlation between the $\gamma$-ray and optical variability of BL~Lacertae during a period of high flare activity over a relatively short time interval (September 1--29, 2022) based on the unprecedentedly high sampling rate of the optical data supports the synchrotron self-Compton mechanism as the origin of the $\gamma$-rays.
The presence of a significant SSC contribution has been established through correlation studies of a large sample of blazars (e.g., \cite{Bhatta21, Jaeger23, Hov14, Liod19}). BL Lacertae has a redshift of $z=0.067$ \cite{Hawley77}, which avoids the difficulties of SSC models that arise at $z\geqslant0.3$, namely the requirement of a compact emission region to account for intraday variability and, consequently, either a high density of radiating electrons leading to internal $\gamma$-$\gamma$ absorption, or an extremely large ($>100$) Doppler factor \cite{Abdo11BLLac}.

The synchrotron self-Compton mechanism is also manifested over a longer time interval (April 1, 2022 -- September 12, 2023), encompassing the last high-amplitude $\gamma$-ray flare during the period of strong BL~Lacertae activity, which lasted from approximately mid-2020 to early 2023.
Based on long-term observational data, we report for the first time a second statistically significant peak in the DCF at the 3$\sigma$ level, corresponding to a time delay of approximately 12 days. 
Previously, Weaver et al. \cite{Weaver20} reported three peaks in the DCF at a $2\sigma$ significance level in the correlation between gamma-ray and optical variability during 2019
September 12 -- October 6, but did not discuss their physical origin.
To identify the causes underlying the emergence of this second peak in the DCF, we performed modeling of light curves in both bands under various assumptions regarding the correlation and time delay of the flares.
For each assumption, 10,000 pairs of light curves were simulated, for which DCFs were computed and the intervals containing 68.2, 95.4 and 99.6\% of the DCF points were determined. Only in two cases (A and B) DCFs exhibit a statistically stable two-peak form.

\textbf{Case A}: The optical flare occurs simultaneously with the second sub-flare in the $\gamma$-rays. The first $\gamma$-ray sub-flare precedes this event by a time interval equal to the separation between the peaks in the DCF. 
This configuration can be realized within the following scenario. Freshly accelerated electrons have been injected into the jet. The jet's magnetic field and electron energies are such that synchrotron radiation is emitted at frequencies significantly higher than those in the optical band. Gamma-ray photons are subsequently generated via inverse Compton scattering of external photons (see, for example, \cite{Dermer93}).
The blob propagates downstream along the jet. 
The first $\gamma$-ray sub-flare ceases as the electrons lose energy through radiative processes, producing ICS/CMB photons at frequencies that continuously decrease.
Moving further along the jet, the magnetic field strength of which decreases with distance \cite{Konigl81}, the synchrotron radiation emitted by the electrons shifts to progressively lower frequencies, eventually reaching the optical band and producing the observed flare.
As a result of SSC, the synchronous flare occurs in the $\gamma$-rays. 
The differing durations and decay rates of the $\gamma$-ray and optical flares can be interpreted by the fact that the main contribution to the observed radiation in both ranges is made by electrons of different energies.
Specifically, the $\gamma$-rays are produced by electrons with energies close to the maximum energy of the power-law spectrum of radiating particles, whereas the synchrotron optical emission is generated by electrons with significantly lower energies.
Thus, the radiative cooling timescale of the electrons, which depends on the particle energy \cite{Pachol}, governs the decay parameters of the flares.

\textbf{Case B}: The $\gamma$-ray flare occurs simultaneously with the first optical sub-flare. After a time interval determined by the separation between the peaks in the DCF, a second optical sub-flare takes place. In the first optical sub-flare, the emission is produced by the synchrotron mechanism, which provides the seed photons for the synchrotron self-Compton emission in the $\gamma$-ray range. The second optical sub-flare arises from ICS of radio photons into the optical band (the possibility of a such process was considered, for example, in \cite{Hutsem10}).
The different time scales of the radiation cooling of electrons producing synchrotron radiation in the optical and radio bands interpret the different duration of the attenuation phase of the first and second optical sub-flares.
In this case, the time delay between optical sub-flares may be caused by the fact that after creating a flare synchronous in two ranges, the blob propagates downstream along the jet, whose medium is optically thick for radio emission at the frequencies required for ICS of photons into the optical band.

The slower decay of the optical flare compared to the gamma-ray flare indicates that long-term variability is dominated by the electron radiative cooling timescale, rather than the light-crossing or perturbation propagation time across the emission region, unlike in some other objects \cite{Chatterjee12}. The discussed scenarios for the two-peaked shape of the DCF may be further complicated by the possible scattering of synchrotron radiation in a nearby cloud, with a subsequent ``return'' into the jet \cite{Aliu14}. Random motions of different parts of the emission region can introduce additional relativistic effects \cite{Narayan12} Furthermore, in the combined SSC and external Compton model, different decay rates of the energy densities of the external radiation and the jet magnetic field can induce delays in the optical variability relative to the $\gamma$-ray variability on the order of a day \cite{Janiak12}.

Thus, interpretation of the shape and position of statistically significant DCF peaks can lead to important conclusions about the physical parameters in the emitting regions and their evolution along the jet.
We emphasize that we do not consider jet bending as a possible mechanism for flare formation, since the studied timescales are significantly shorter than those required for flare production via jet curvature.
Modeling has shown that the results remain robust even with a substantial reduction in the number of data points in both model time series. 
This fact gives the proposed method an advantage over direct light-curve analysis in cases of sparse or irregular sampling. 
However, the interpretation of DCF can be ambiguous. Therefore, simultaneous observations in the radio and X-ray bands are necessary to constrain the physical processes operating in the object.

\section*{Acknowledgments}

This research has made use of the SVO Filter Profile Service "Carlos Rodrigo", funded by MCIN/AEI/ 10.13039/ 501100011033/ through grant PID2023-146210NB-I00.
We used Fermi-LAT data through Monitored Source List Light Curves (https://fermi.gsfc.nasa.gov/ssc/data/access/lat/msl\_lc/) provided by the LAT team.
Moreover, this paper includes data collected by the TESS missions and obtained from the MAST data archive at the Space Telescope Science Institute (STScI). Funding for the TESS mission is provided by the NASA Explorer Program.
We gratefully acknowledge the contributions of the AAVSO observer community, whose photometric data and metadata resources were used in this study and made available through the AAVSO's scientific archives.

\bibliographystyle{aasjournal}

\bibliography{oja_template}

\begin{appendix}

\section{Light curve linear shift}
\label{ap:app1}

Let us consider two data series: in the g and R bands of the AB and Johnson-Cousins photometric systems. According to the formulae relating magnitude and spectral flux density $F$, the difference in magnitudes between the two photometric systems is given by
\begin{equation}
    \label{eq:mag-dif}
    m_\text{AB}-m_\text{R}=-2.5\log\!\left(\frac{F_\text{g}}{F_\text{R}}\right)+0.18.
\end{equation}
Accounting the power-law spectrum of blazar optical emission ($F\sim\nu^{-\alpha}$, where $\alpha$ is a spectral index) and assuming that the detected radiation corresponds to the effective frequency of each filter, we obtain the argument of the logarithm is $(\nu_\text{R}/\nu_\text{g})^\alpha$ (where $\nu_\text{g\,(R)}$ is the effective frequency of g (R) band). 
If the spectral index remains unchanged as the object's brightness varies, the difference between the magnitudes measured in two filters remains constant. In that case, when one data series is aligned with another by a linear shift, minimizing the residual between the linear fit to the points of the primary series and the measurements of the series being aligned, an error can arise in the values of the aligned series. 
This error is caused by the fact that the true g-band magnitude at some time $t$, corresponding to the time of the R-band observation, is unknown to the observer and may differ from the value given by the linear fit through the two neighboring points in g-light curve measured at times $t_1< t$ and $t_2 > t$.
The variability of blazars, including BL Lacertae, exhibits both a constant spectral index $\alpha$ as the source brightens and a decreasing 
$\alpha$, known as the bluer-when-brighter (BWB) trend \cite{Gorbachev24b, LiGuo24}. This constitutes a second reason for the potential discrepancy between the true g-band magnitude and the value recovered from R-band data.

To estimate the magnitude uncertainty in a combined light curve constructed from two data series, we performed numerical simulations.

(i) During the variability of the spectral flux $F$, its power spectral density follows a power law with a negative slope \cite{TK95}. 
Following the method described in \cite{TK95}, we generated 5,000 light curves for $F$, with the power spectral density slope for each curve drawn randomly between $-2$ and $-1$.
The simulated light curves have a temporal resolution of 0.5 days and a duration of 512 days, which approximately corresponds to the time span of the g-band data series. We set these model data as g-band series.

\begin{figure}[!h] %
	\centering
	\includegraphics[width=.8\textwidth]{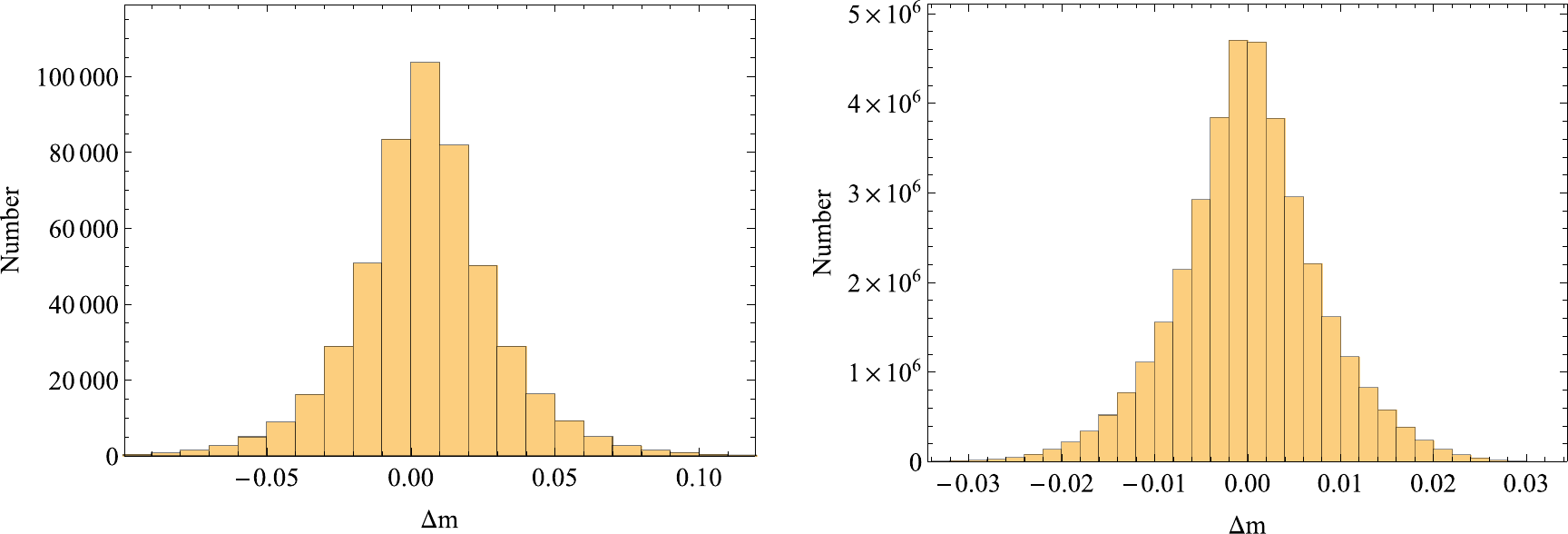}
	\caption{Distribution of deviations: a) of the second series in the g band relative to its initial position; b) of the true g-band values of data points from those recovered by linearly shifting the R-band light curve.}
	\label{figA1}
\end{figure}

After converting the flux values to magnitudes, we extracted two series of randomly selected points from each model light curve. The number of points was randomly drawn from the ranges 100$-$150 and 50$-$100 for the first and second series, respectively. Then, the magnitude values of the second series were increased by 1 and then gradually decreased until the sum of squared differences between the magnitudes of second series and the linear interpolations of the first series, evaluated at the times of the second series, reached a minimum. We repeated this procedure 100 times for each initially modeled light curve. The obtained differences between the obtained shift of the second series and its initial offset is significantly smaller than the variability amplitude (Fig.~\ref{figA1}a).

(ii) To construct a light curve corresponding to another effective frequency, such as the R-band frequency, while accounting for the occasionally observed BWB trend \cite{Gorbachev24b, LiGuo24}, we relied on the observational fact that data points from simultaneous multi-band observations are well approximated by a straight line on a flux-flux diagram \cite{Hagen97,Larionov08}. In this diagram, straight lines whose points satisfy the condition $\alpha=\text{const}$ exit the origin at different angles for different values of $\alpha$. A straight line fitted to the observational data intersects the lines with  $\alpha=\text{const}$ if the data exhibit changes in $\alpha$ during flux variability \cite{Gorbachev2022, Weaver20}. Therefore, we set $\alpha_1$ as the realization of a random variable in the range of 1 to 2.2, consistent with observations \cite{Gorbachev24b, LiGuo24}, and associated this value with the minimum flux in the g band $F_{g,\,1}=10$~mJy. To simulate a randomized BWB trend, we assign $\alpha_2$ corresponding to the maximum flux $F_{g,\,2}=100$~mJy and its value is given by $\alpha_2=\alpha_1-\Delta \alpha$, where $\Delta \alpha$ is the realization of a random variable drawn from the range 0$-$0.5.
Thus, the R-band flux was defined by the equation
\begin{equation}
    \label{eq:FR_lin}
    F_\text{R}=A F_\text{g}+B+\sigma,
\end{equation}
where 
$$A=\frac{F_{\text{R},\,2}-F_{\text{R},\,1}}{F_{\text{g},\,2}-F_{\text{g},\,1}},$$
$$F_{\text{R},\,1 (2)}=\left(\frac{\nu_\text{g}}{\nu_\text{R}}\right)^{\alpha_{1,(2)}},$$
$\sigma$ characterizes the scatter of the data points and is a realization of a normally distributed random variable with a mean of 0 and a standard deviation of 5~mJy.

(iii) To generate a sparse, unevenly sampled time series from dense, evenly sampled data series that more closely resembles actual observations, we performed 100 iterations for each pair of light curves, randomly selecting points to construct the final pair of light curves. The number of points was drawn randomly from the range 100 to 150 for the data simulating the  g-band series, and from 50 to 100 for the model R-band series.

(iv) After converting the flux values to magnitudes, we applied the linear shift of the R-band series relative to the g-band series, as described in Section~2. Next, we calculated the deviations of the resulting shifted R-band values from the originally simulated g-band points at the corresponding epochs. It is seen that the deviations are significantly smaller than the variability amplitude (Fig.~\ref{figA1}b).

\section{Significance level estimation}

\begin{figure}[!h] %
	\centering
	\includegraphics[width=.8\textwidth]{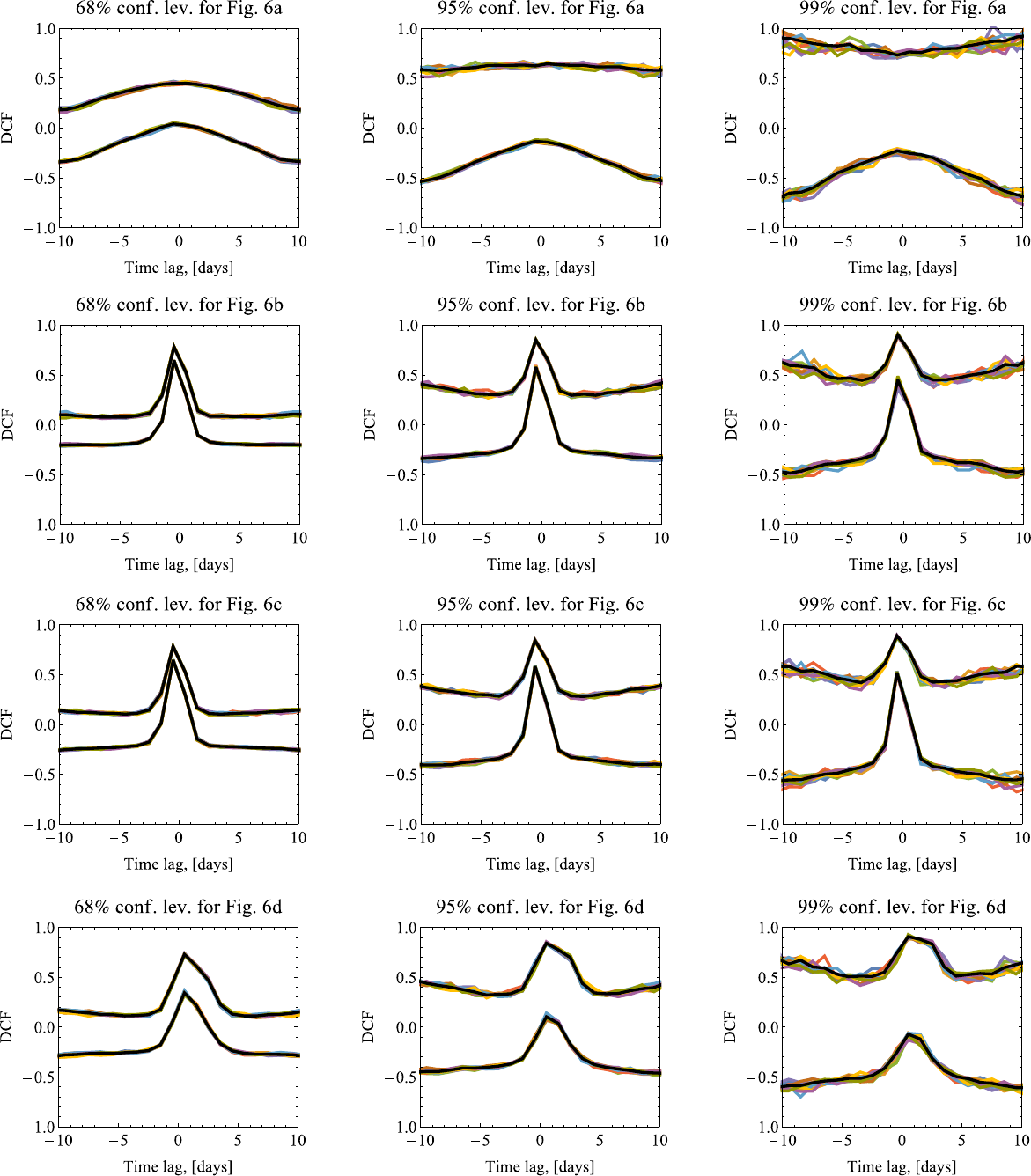}
	\caption{ Confidence intervals containing 68.2\%, 95.4\%, and 99.6\% of the DCF points, calculated under different assumptions corresponding to cases (a), (b), (c), and (e) in Section~\ref{lcm}. The colored and black lines mark confidence levels obtained for ten sets of 1,000 simulated DCFs and the entire modeled dataset, respectively.}
	\label{A2}
\end{figure}

\begin{figure}[!h] %
	\centering
	\includegraphics[width=.8\textwidth]{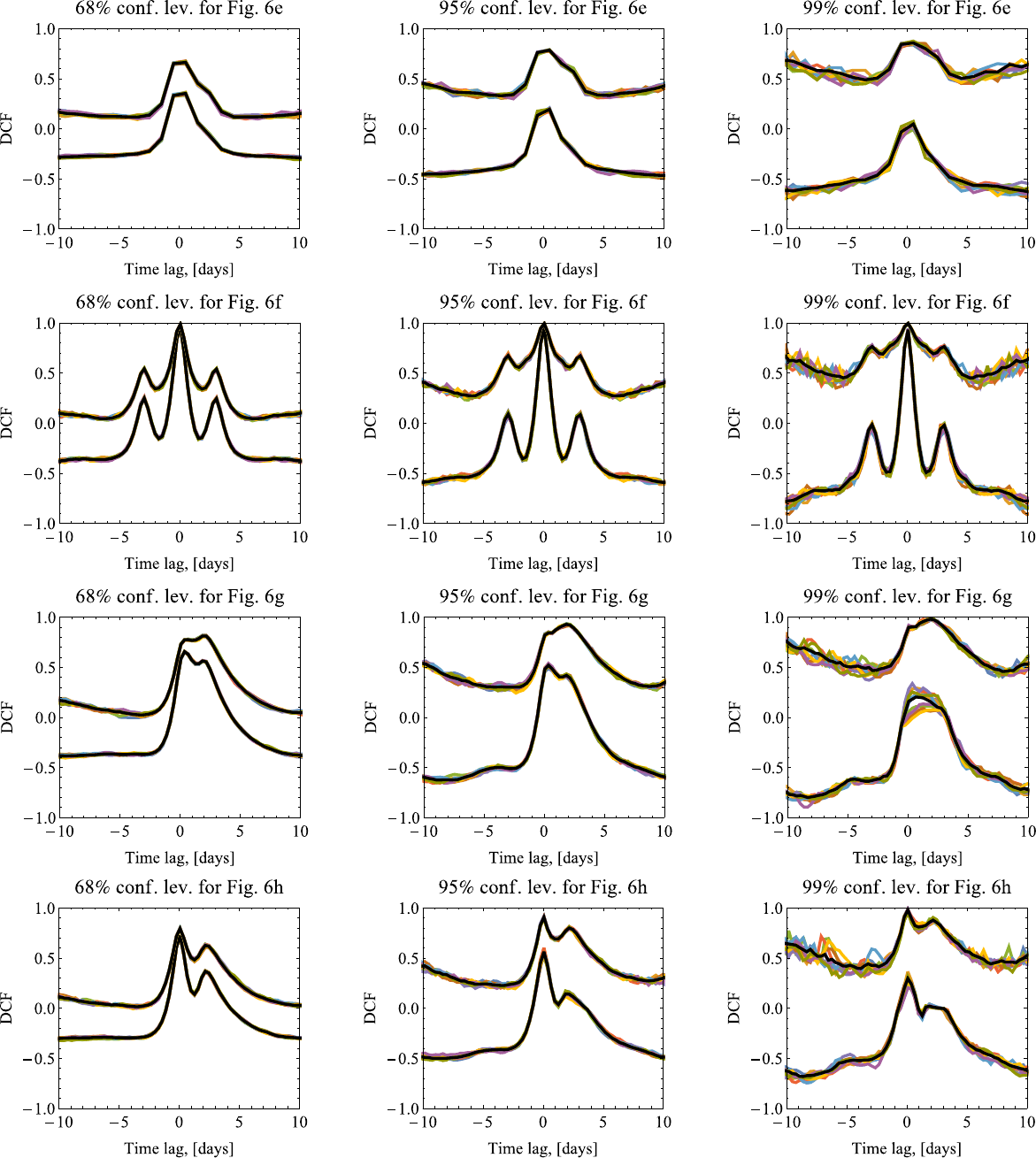}
	\caption{Confidence intervals containing 68.2\%, 95.4\%, and 99.6\% of the DCF points, calculated under different assumptions corresponding to cases (f), (g), (h), and (f) in Section~\ref{lcm}. The colored and black lines mark confidence levels obtained for ten sets of 1,000 simulated DCFs and the entire modeled dataset, respectively.}
	\label{A3}
\end{figure}

To answer the question of how many simulations are enough to adequately estimate the significance levels of the DCF, we compared the shapes of 68.2\%, 95.4\%, and 99.6\% confidence intervals calculated for 10 sets of 1,000 simulated pairs of light curves, as well as for the entire dataset. We performed modeling based on the assumptions described in Section~\ref{lcm}. The result, presented in Figs.~\ref{A2} and \ref{A3}, shows that 1,000 simulations are sufficient for a robust estimation of 68.2\% and 95.4\% significance levels. 
A tenfold increase in the number of model light curve pairs leads to the smoothing of the ``wobbling'' shape for the 99.6\% significance level interval.

\end{appendix}

\end{document}